# Two-photon photoactivated voltage imaging in tissue with an Archaerhodopsin-derived reporter

Miao-Ping Chien*[1], Daan Brinks*[1], Yoav Adam[1], William Bloxham[1], Simon Kheifets[1], Adam E. Cohen[1,2]†

[1] Departments of Chemistry and Chemical Biology, Harvard University, Cambridge, MA 02138, USA

[2] Howard Hughes Medical Institute, Cambridge, MA 02138, USA.

†E-mail: cohen@chemistry.harvard.edu

* Joint first author

## Abstract
Robust voltage imaging in tissue remains a technical challenge. Existing combinations of genetically encoded voltage indicators (GEVIs) and microscopy techniques cannot simultaneously achieve sufficiently high voltage sensitivity, background rejection, and time resolution for high-resolution mapping of sub-cellular voltage dynamics in intact brain tissue. We developed a pooled high-throughput screening approach to identify Archaerhodopsin mutants with unusual photophysical properties. After screening ~$10^5$ cells, we identified a novel GEVI, NovArch, whose 1-photon near infrared fluorescence is reversibly enhanced by weak 2-photon excitation. Because the 2-photon excitation acts catalytically rather than stoichiometrically, high fluorescence signals, optical sectioning, and high time resolution are achieved simultaneously, at modest 2-photon laser power. We developed a microscopy system optimized for NovArch imaging in tissue. The combination of protein and optical engineering enhanced signal contrast sufficiently to enable optical mapping of back-propagating action potentials in dendrites in acute mouse brain slice.

## Introduction

Improved tools for voltage imaging in tissue would be a powerful enabling capability across broad domains of neuroscience[1-3]. Recent advances in genetically encoded fluorescent voltage indicators (GEVIs) have enabled all-optical electrophysiology (Optopatch) in a transgenic mouse[4], and optical mapping of spontaneous and sensory evoked activity in mouse brain *in vivo*[5]. Nonetheless, background autofluorescence and optical scattering remain obstacles to achieving deep, sub-cellular, or wide-field voltage imaging in tissue. Better reporters and better microscopes are needed.

Scanning two-photon (2P) microscopy has revolutionized $Ca^{2+}$ imaging *in vivo*, enabling mapping of neuronal excitation in many species, brain regions, and stimulus paradigms [6, 7]. A natural impulse is to apply the same technique to voltage imaging. Recent progress in 2P imaging of a voltage-sensitive dye shows promise, but the dye lacks cell-type selectivity [8]. 2P imaging of the GEVI ASAP2s has been achieved in drosophila brain and organotypic slice, but not yet in acute rodent brain tissue where optical scattering is much more severe [9]. We previously explored 2P microscopy of a panel of GEVIs [10]. The QuasAr GEVIs, which are advantageous from the perspective of sensitivity, speed, and spectral compatibility with channelrhodopsin, had far too little 2P fluorescence to be useful. Other GEVIs gave detectable 2P signals, but suffered from rapid photobleaching. These challenges trace to fundamental differences between voltage and Ca$^{2+}$ imaging, namely: (a) neuronal action potentials are typically 100 – 1000-fold faster than fluorescence transients of Ca$^{2+}$ reporters; (b) voltage signals come only from the 2-dimensional cell membrane while Ca$^{2+}$ signals come from the 3-dimensional bulk cytosol; and (c) diffusive exchange of photobleached reporter molecules for fresh ones is ~100-fold slower in membranes than in cytosol due to the correspondingly lower diffusion coefficients of transmembrane vs. cytosolic proteins. Together, these factors imply that in 2P voltage imaging as compared to Ca$^{2+}$ imaging, a much greater flux of photons must come from a much smaller number of molecules, hence the difficulty in achieving simultaneously high time resolution, voltage sensitivity, and robustness to photobleaching. Quantitative estimates by us[10] and others[11] suggest that large-area 2P voltage imaging in intact tissue will be a technically difficult route to pursue.

In view of these constraints, how can one achieve robust voltage imaging in tissue? We explored whether the complex photocycles of microbial rhodopsins could open the door to nontraditional imaging techniques. Microbial rhodopsins have multiple conformational states whose interconversion is driven by light and/or voltage[12]. One can further engineer or evolve these proteins to have specific photocycle topologies connecting fluorescent states and voltage-sensitive states. We previously engineered these multi-state dynamics to develop microbial rhodopsin reporters for absolute voltage[13], light-gated voltage integrators, and a light-gated voltage sample-and-hold[14].

In wild-type Archaerhodospin 3 (Arch), the voltage-sensitive fluorescence arose through a sequential three-photon process[12]. This unusual attribute in principle allowed optical sectioning, because the fluorescence showed a nonlinear enhancement at regions of high illumination intensity. However, the effect saturated at illumination intensities too low to be practically useful, and in wild-type Arch the light-induced photocurrent perturbed the membrane voltage. This nonlinearity was eliminated in the protein engineering efforts that led to the QuasAr GEVIs[15]. We recently developed an improved QuasAr GEVI, called QuasAr3, which showed enhanced expression and trafficking *in vivo* compared to the previously published QuasAr2 [16]. Introduction of the mutation V59A on the QuasAr3 background led to another GEVI, termed photoactivated QuasAr3 (paQuasAr3), which had the unusual property that its near infrared fluorescence was reversibly enhanced by weak blue illumination [16]. The ~2-fold photoactivation of paQuasAr3 was beneficial for voltage imaging of near-surface neurons *in vivo*, but was insufficient to enable optically sectioned imaging at greater depth.

Here we evolve paQuasAr3 into a non-pumping Arch mutant, NovArch, in which fluorescence nonlinearities occur with parameters useful for imaging in tissue. To apply NovArch in tissue, we develop a novel mixed-wavelength three-photon excitation microscope. We combine the novel sensor and the novel microscope to record action potentials to a subsurface depth of 220 μm and to image back-propagating action potentials in dendrites in acute mouse brain slice.

**Results**
**Two-photon reversibly photoactivatable QuasAr for optical sectioning**

We began with a detailed photophysical characterization of paQuasAr3. In HEK 293T cells expressing paQuasAr3, red illumination (λ = 640 nm, 90 W/cm$^2$) induced weak near infrared fluorescence ($F_R$). Addition of much dimmer blue light (λ = 488 nm, 12 W/cm$^2$) on top of the red illumination sharply increased the fluorescence (Fig. 1A). Initially we assumed that the blue light was directly exciting fluorescence of the GEVI. However, rapidly extinguishing the blue illumination while maintaining the red illumination led to a gradual decrease of the enhanced fluorescence, with a time constant of ~450 ms (Fig. 1A). Furthermore, the fluorescence from simultaneous red and blue illumination ($F_{R,B}$) was on average more than twice the sum of the fluorescence from red-only ($F_R$) and blue-only ($F_B$) illumination (Fig. 1B) ($F_B$ may partially have come from Citrine fluorescence leakage through the emission filter). We quantified this excess non-additive fluorescence by ($F_{R,B}$-$F_B$-$F_R$)/$F_R$ = $\Delta F^{blue}/F_R$ and found $\Delta F^{blue}/F_R$ = 1.3 ± 0.5 (mean ± s.d., $n$ = 7 cells). The parent protein, QuasAr3, was previously reported to show only very slight (~1%) photoactivation under blue light illumination[16].

Intrigued by this photoactivation effect, we then tested whether the blue activation photon could be replaced by two infrared photons, coherently absorbed from a femtosecond laser pulse. Figure 1C shows that two-photon (2P) excitation ($\lambda$ = 900 nm, 3 mW) scanned along the periphery of a HEK cell expressing paQuasAr3 reversibly enhanced the red-excited near infrared emission, while 2P illumination alone did not induce detectable near infrared fluorescence. We defined ($F_{R,2P}$-$F_{2P}$-$F_R$)/$F_R$ = $\Delta F^{2P}/F_R$ and found $\Delta F^{2P}/F_R$ = 0.4 ± 0.1 (mean ± s.d. $n$ = 14, $P_{900nm}$ = 3 mW). We ascribe the smaller 2P enhancement compared to the 1P blue light enhancement to the fact that the 2P scan pattern only covered a portion of the cell membrane. QuasAr3 did not show detectable enhanced fluorescence under 2P infrared excitation (Fig. S1).

We then quantified the effect of 2P illumination on the voltage-sensitive fluorescence of paQuasAr3. Fluorescence under red light illumination alone showed a voltage sensitivity of $\Delta F_R^{(V)}/F_R$ = 23 ± 6% for a voltage increase from -70 to +30 mV (mean ± s.d., $n$ = 4 cells, Fig. 1D). Addition of 2P illumination increased both the baseline fluorescence and the voltage-induced changes in fluorescence by the same factor of $\Delta F^{2P}/F_R$ = 0.4, so the fractional voltage sensitivity remained the same: $\Delta F_{R,2P}^V/F_{R,2P}$ = 26 ± 6% for a voltage increase from -70 to +30 mV (mean ± s.d., $n$ = 4 cells, Fig. 1D). The fluorescence responded to a step in membrane voltage with a fast time constant of 1.6 ms (72%) and a slow time constant of 31 ms (28%; Fig. 1E). 2P illumination did not affect the kinetics of the voltage response (Fig. 1E).

These unusual observations suggested the photocycle model shown in Fig. 1F. Blue illumination converted a non-fluorescent state, $D_1$, into a manifold of states, $\{D_2 \rightleftharpoons F\}$, that showed voltage-sensitive one-photon fluorescence. Cessation of blue illumination led to a gradual reversion to the non-fluorescent $D_1$ state. The blue photon could be replaced by two coherently absorbed near-infrared photons. While the underlying photocycle is likely more complex, this model captures the main features of the data. The key aspect for what follows is that the sensitizing light acts catalytically, rather than stoichiometrically, in governing fluorescence: absorption of a single blue photon or a pair of infrared photons can lead to many cycles of red excitation and near infrared fluorescence. The photocycle of Fig. 1F suggested that it might be possible to perform optically sectioned voltage imaging in tissue by intersecting a 1P blue or 2P infrared photoactivation beam with a 1P red imaging beam.

**Screening for improved photoactivatable GEVIs**
To achieve optically sectioned voltage imaging in tissue, we first had to improve the photoactivation ratio beyond the $\Delta F^{blue}/F_R$ = 1.3 observed for 1P activation of paQuasAr3. We used a novel screening technique, termed Photostick, to select single cells optically from a pooled library culture, using photoactivation as the selection criterion (Figure 2A)[17]. Starting with paQuasAr3, we made a library of mutants using error-prone PCR at a mean mutation rate of 2.8 mutations per gene. This pooled library was introduced into a lentiviral vector. HEK cells were infected with this library at low multiplicity of infection (~$10^{-3}$) to ensure that few cells were infected with two viral particles. The expressing cells were selected with puromycin and plated at a density of ~50k/cm$^2$ on coverslip-bottomed dishes.

The cells were then imaged on a custom-built ultrawide-field microscope equipped with a low magnification (2x) high numerical aperture (NA 0.5) objective (Fig. S2), which yielded a field of view 1.2 mm × 3.3 mm, comprising nominally ~$2\times10^3$ cells. Near infrared fluorescence of each cell was quantified under red only ($F_R$), blue only ($F_B$) and simultaneous red and blue ($F_{R,B}$) illumination. The *Materials and Methods* contain a detailed discussion of the illumination protocol, the calibration of the red and blue illumination profiles, and the image processing steps. Our figure of merit was the blue light-induced nonlinear enhancement in fluorescence, $\Delta F^{blue}/F_R$ (Figs. 2B,C). We screened 57 fields of view, corresponding to ~$10^5$ cells. Cells with high brightness and large fractional enhancement in fluorescence upon blue illumination were judged to be hits (Fig. 2D).

Hits were isolated via photochemical crosslinking to the dish. A photochemical crosslinker, 4-fluoro-3-nitrophenyl azide (FNPA) (15 µM final concentration), was added to the imaging medium. A pair of galvo mirrors directed light from a 405 nm laser (15 mW) sequentially to each selected cell for ~200 ms (*Methods*). The crosslinker immobilized the target cells to the dish. Non-target cells were removed via a wash with trypsin (Fig. 2E). Target cells were then aspirated into ~10 µL of phosphate buffered saline (PBS) and dispersed for single-cell PCR. The QuasAr mutant gene was amplified from each target, sequenced, and re-cloned for validation and testing. The hit with greatest photo-activation contained two additional mutations on the paQuasAr3 background: V209I and I213T (Fig. S3). We dubbed paQuasAr3(V209I, I213T) "NovArch" for its rapid increase in brightness upon blue illumination.

**Photophysical characterization of NovArch**
We expressed NovArch in HEK cells and measured the near infrared fluorescence under red illumination ($\lambda$ = 640 nm, 90 W/cm$^2$) with blue pulses superposed at variable intensity. We observed an up to 8-fold enhancement in near infrared fluorescence upon saturating blue illumination ($\Delta F^{blue}/F_R$ = 7, Fig. 3A), with 50% maximum enhancement at a blue intensity of $I_{blue}$ = 3.9 W/cm$^2$. For blue illumination intensity compatible with tissue imaging, $I_{488nm}$ = 12 W/cm$^2$, the blue illumination-induced nonlinear enhancement in fluorescence was $\Delta F^{blue}/F_R$ = 4.8 ± 0.9 (mean ± s.d., $n$ = 7, Fig. 3B). We cloned additional constructs in which each novel mutation was individually reversed and tested the nonlinear photoactivation. None of these constructs had photoactivation as large as NovArch (Fig. S4A), demonstrating that both novel mutations were important.

Upon onset of the blue illumination, NovArch fluorescence rose with single-exponential kinetics and time constant inversely related to illumination intensity: at $I_{blue}$ = 6 W/cm$^2$ we measured $\tau_{up}$ = 14 ms, and at $I_{blue}$ = 100 W/cm$^2$, $\tau_{up}$ = 4 ms (Fig. 3A, bottom). At the lower blue intensity (6 W/cm$^2$) this time constant implies an absorption cross section for the photoactivation process of $\sigma$ = $4.8\times10^{-18}$ cm$^2$, corresponding to an extinction coefficient $\varepsilon$ = 1270 M$^{-1}$cm$^{-1}$. Upon cessation of the blue illumination, the fluorescence enhancement decayed with double-exponential kinetics, comprising a fast component ($\tau_1 = 74$ ms, 78%) and a slow component ($\tau_2 = 2.6$ s, 22%, Fig. 3A).

We next tested the 2P activation of NovArch. HEK cells expressing NovArch were exposed to wide-field red illumination as above, and illumination from an ultrafast 2P pulsed laser ($\lambda = 900$ nm) was scanned around the periphery of the cell at 500 Hz. The 2P illumination reversibly enhanced the NovArch fluorescence 2.7-fold (Fig. 3C), corresponding to $\Delta F^{2P}/F_R = 1.7 \pm 0.4$ (mean ± s.d., $n = 12$). NovArch retained good voltage sensitivity (41 ± 7% increase from -70 to +30 mV; Fig. 3E) and speed of response (1.2 ms, 76%; 10 ms, 24%; Fig. 3F), with similar sensitivity and kinetics with and without 2P sensitization.

To determine the optimal wavelengths for NovArch imaging and activation we measured the fluorescence excitation and photoactivation spectra. We expressed NovArch in HEK cells, centrifuged the cells to form a dense pellet, and performed spectroscopy in a home-built microscope system. Using light from a tunable supercontinuum laser (90 µW per band) we measured the direct fluorescence excitation spectrum, which peaked at 620 nm, validating the choice of 640 nm light for fluorescence excitation (*Methods*). We combined the tunable supercontinuum beam with a blue beam (488 nm, 1 mW) and measured the fluorescence excitation spectrum of the photoactivated state. This spectrum had a similar shape, but higher amplitude, compared to the unactivated state (Fig. 3G), confirming that the fluorescence was dominated by a single state whose population was modulated by the blue illumination, as in the photocycle model (Fig. 1F). We then combined the tunable supercontinuum beam with a red ($\lambda = 635$ nm, 1 mW) beam and measured the photoactivation action spectrum, which peaked at 470 nm (Fig. 3H), validating the choice of 488 nm as the activation wavelength.

Finally, we measured the effect of 2P illumination parameters on NovArch photoactivation, with the goal of learning how best to activate NovArch fluorescence in tissue. 2P activation was more efficient at $\lambda = 900$ nm than at 1,000 nm or 1,100 nm (Fig 3I). 2P photoactivation showed saturation behavior, with 50% maximum activation at a 2P laser power of 1.8 mW (Fig. S4B). Upon onset of the 2P illumination, the fluorescence rose with single-exponential kinetics and time constant inversely related to illumination power: at $P_{2P} = 2$ mW we measured $\tau_{up} = 160$ ms, and at $P_{2P} = 12$ mW, $\tau_{up} = 40$ ms. Upon cessation of the two-photon illumination, the fluorescence enhancement decayed with double-exponential kinetics, comprising a fast component ($\tau_1 = 165$ ms, 19%) and a slow component ($\tau_2 = 2.4$ s, 81%).

We tested NovArch expression and its ability to report action potentials in cultured rat hippocampal neurons. NovArch showed excellent trafficking in the soma and throughout the dendritic arbor (Fig. 4A). Under wide-field red illumination (110 W/cm$^2$) action potentials were detected in single-trial recordings with an amplitude of $\Delta F^V/F_R = 28 \pm 5\%$ (mean ± s.d., $n = 6$; Fig. 4B). The fluorescence recordings showed close correspondence to simultaneous manual patch clamp recordings. Addition of 2P illumination (900 nm, 3 mW) scanned around the cell periphery led to a 2.8-fold increase in both the baseline fluorescence and the spike amplitudes, i.e. $\Delta F^{2P}/F_R = 1.8 \pm 0.4$ (mean ± s.d., $n = 6$; Fig. 4B). Compared to the baseline fluorescence in the absence of 2P activation, $F_R$, the 2P-activated spikes had an amplitude of $\Delta F^{V,2P}/F_R = 79 \pm 11\%$ (mean ± s.d., $n = 6$). The parent protein, QuasAr3, also reported spikes in cultured neurons, but did not show 2P enhancement in spike amplitude (Fig. 4C).

**NovArch voltage imaging in tissue**
We next sought to apply NovArch in tissue, using sparsely expressed Cre-dependent constructs for paQuasAr3-Citrine and NovArch-Citrine, delivered via AAV2/9 (*Methods*). The primary obstacles to voltage imaging in tissue are scattering of excitation and emission light, and the high background from out-of-focus fluorescence. We explored several optical approaches to maximize the signal-to-background ratio (SBR) for NovArch in tissue.

Voltage-sensitive fluorescence comes only from the cell membranes, a two-dimensional manifold embedded in a three-dimensional medium. Any light that enters a tissue and misses the cell membrane contributes solely to background, and not to useful signal. Thus one should focus the illumination on the cell membrane only.

Figure 5 shows the impact of successively more precise optical targeting of excitation in acute brain slice. In each case we first acquired a conventional 2P fluorescence image of the appended Citrine tag to provide ground truth on cell morphology and to guide the targeted illumination (Fig. 5A). For measurements of near infrared GEVI fluorescence, we collected images on a back-illuminated EMCCD camera (Andor Ixon3 860).

Wide-field illumination of a cell 20 μm below the surface produced a low contrast image, with a SBR of only 0.19 (Fig 5B). Wide-field illumination was clearly ill-suited to voltage imaging in tissue because many photons missed the target cell altogether, yet still contributed to background fluorescence. Closing a field aperture to restrict illumination just to the soma of the targeted cell improved the SBR more than 8-fold, to 1.55 (Fig. 5C).

Not all parts of the cell membrane contribute equally to the signal: redirecting an excitation photon from a dim region to a bright region on the membrane enhances the likelihood of the photon producing useful fluorescence, without changing the likelihood of the photon contributing to background. Geometrical projection effects cause the cell membrane to appear ~3-fold brighter at the equatorial periphery than at the polar caps, so excitation is ~3-fold more efficient when targeted to the equatorial periphery. We used a pair of fast galvo mirrors to scan a red (640 nm) laser focus around the periphery. This 'semi-confocal' approach (focal excitation, wide-field detection) further improved the SBR to 3.5, a 19-fold improvement over wide-field illumination (Fig. 5D). This enhancement is applicable to all GEVIs, and does not require the photoactivation property of NovArch.

We then sought to image cells deeper in the slice. At a depth of 120 μm, wide-field illumination produced no detectable contrast. Cell-localized excitation produced a faint blur in the cell location. Scanning focal excitation and wide-field detection revealed a faint outline of the cell shape. We then co-aligned the 2P laser focus (900 nm, 6.8 mW) with the 1P laser focus (637 nm, 1.7 mW; Fig. S5), and jointly scanned the pair of lasers around the cell periphery. For paQuasAr3, this approach did not produce detectable enhancement, presumably due to the small photoactivation $\Delta F^{opt}/F_0$ for paQuasAr3. For NovArch, the 2P enhancement clearly revealed the outline of the cell (Fig. 5E). The 2P enhancement of the NovArch signal closely matched the conventional 2P image of the Citrine expression marker (Fig. 5F). These results demonstrate that 2P-activated 1P

NovArch imaging can resolve single cells at depths where conventional 1P excitation wide-field or confocal approaches fail (Fig 5G).

We next applied 2P-enhanced NovArch imaging to study voltage signals in acute slice. Using 2P Citrine images to define the cell periphery, we co-scanned 2P (900 nm, 6.8 mW) and 1P (637 nm, 1.7 mW) foci around cells at depths between 45 and 220 μm (Fig. 6A). The period of the galvo cycles was synchronized with the frame rate of the camera, both to 500 Hz. Fluorescence signals were low-pass filtered to an effective frame-rate of 370 Hz. 2P sensitization enhanced the spike amplitude by a factor of $3.5 \pm 0.5$ (mean ± s.e.m, $n = 14$ cells between 40 and 70 μm into slice). The 2P-sensitized SNR was $96 \pm 12$ (mean ± s.e.m, $n = 14$ cells between 40 and 70 μm into slice; Fig 6B). The SNR of single spikes diminished approximately exponentially with depth. The decay length was ~150 μm both without and with 2P sensitization, and prefactors were 39 (without 2P sensitization) and 128 (with 2P sensitization; $n = 28$ cells; Fig 6C). With red-only semi-confocal illumination, single-trial action potentials were resolved to a depth of 100 μm. With 2P sensitization, single-trial action potentials were resolved to a depth of 220 μm. The combination of membrane-localized excitation and 2P NovArch sensitization dramatically extends the depth to which voltage imaging can be performed in tissue.

**Combining NovArch imaging with optogenetic stimulation**
A key merit of near-infrared GEVIs is that they are, in principle, compatible with optogenetic stimulation, for all-optical electrophysiology. We thus sought to combine NovArch imaging with optogenetic stimulation of a blue-shifted channelrhodopsin, CheRiff. The complex photophysics of NovArch raised the possibility of two types of optical crosstalk: (1) the 2P beam used to sensitize NovArch could also activate CheRiff, leading to spurious activation of the target cell; and (2) the blue light used to activate CheRiff could further sensitize NovArch, leading to spurious signals in the NovArch fluorescence channel.

We co-expressed NovArch and a blue-shifted channelrhodopsin, CheRiff, in acute slices (*Methods*), and used wide-field blue stimulation (488 nm, 60 mW/cm$^2$) to activate the CheRiff. Figure 6D shows that tonic blue CheRiff stimulation induced high-frequency spiking, but had only a small effect on the NovArch baseline. 2P illumination enhanced the NovArch fluorescence signal by a factor of 3.5, and did not induce spurious spontaneous activity.

The absence of crosstalk in this system can be explained by the differing photophysical properties of NovArch and CheRiff. To maximize the fluorescence SNR, the 2P sensitization and 1P imaging beams were confined to the brightest part of the cell, the equatorial periphery. While the 2P beam likely excited CheRiff molecules in its path, the cumulative conductance of these molecules was too low to induce spiking. Indeed, due to the low unit conductance of channelrhodopsins, 2P optogenetic stimulation has remained technically challenging, and is best achieved either by spiral scanning [18] or by advanced beam shaping approaches that blanket the cell membrane [19, 20]. The localized scan pattern which favored voltage imaging disfavored 2P CheRiff stimulation.

The absence of crosstalk from blue light into the NovArch fluorescence channel can be explained by the differing sensitivities of CheRiff and NovArch (Fig. 6E). CheRiff has 50% activation at a blue intensity of 20 mW/cm$^2$,[21] while NovArch has 50% of maximum sensitization at a blue intensity of 3.9 W/cm$^2$. At the blue illumination used in these experiments, 60 mW/cm$^2$, the blue light-induced change in NovArch fluorescence is predicted to be < 10%. Thus despite the spectral overlap between CheRiff activation and NovArch sensitization, these two processes can be performed in tissue with little crosstalk.

**Mapping back-propagating action potentials in tissue**
Finally, we explored the capability to map dendritic voltage associated with back-propagating action potentials. We expressed NovArch sparsely *in vivo* and simultaneously expressed CheRiff broadly under the pan-neuronal hSyn promoter. In an acute brain slice, we identified a NovArch expressing L5 pyramidal cell and we mapped the dendritic tree using conventional two-photon microscopy of the appended Citrine fluorescent tag (Figs.7A, S6). We then designed laser scan paths that co-scanned red and 2P foci along sections of dendrite at a series of distances from the cell body. To enhance the mean spike rate of the cell, wide-field blue illumination was delivered in a pattern of 500 ms on, 500 ms off. Along the distal apical dendrite, back-propagating action potentials were clearly visible in single trial to a distance 480 μm from the soma (Fig 7B). Averaging of the recorded spikes showed the broadening and dampening of the back-propagating action potentials as the distance from the soma increased (Fig 7C). These results demonstrate the possibility of mapping sub-cellular voltage dynamics in genetically specified neurons in tissue.

**Discussion**
By combining protein engineering, detailed spectroscopy, and optical design we have demonstrated a novel approach to high-sensitivity voltage imaging in tissue. The NovArch protein functions as a reversibly photo-activated GEVI, enabling 2P optical 'highlighting' of cells or sub-cellular regions of interest. Key to achieving the benefits of NovArch was an optical design which localized the fluorescence excitation to the peripheral cell membrane. This approach maximized the ratio of signal fluorescence to input light, achieving a 19-fold enhancement in signal-to-background ratio compared to wide-field excitation. 2P sensitization of NovArch further enhanced signal levels by an additional factor of 3.5, without increasing background. Together the protein and optical engineering enhanced signal-to-background ratios > 60-fold compared to wide-field epifluorescence imaging of non-photoactivated Arch-derived GEVIs.

A combined 2P and 1P focal scanning approach offered several advantages for voltage imaging. Due to the catalytic nature of the two-photon excitation, the two-photon excitation did not need to be exceptionally bright. Due to the two-photon definition of the sensitized region, the one-photon focus did not need to be exceptionally sharp, so this approach was robust to some degree of light scatter. Due to the comparatively slow relaxation of the NovArch sensitized manifold (~450 ms), the two-photon scanning did not need to be exceptionally fast. In the present implementation, the 2P and 1P beams were co-aligned, requiring fast scanning of both. To image multiple cell in parallel, one can conceive of an optical system where the red light is targeted to multiple cell membranes in a static pattern, e.g. via a diffractive spatial light modulator, while the 2P sensitizing beam visits the target cells serially at a modest repetition rate.

It is instructive to compare the effects of photoactivation to other approaches one might take to improve the SNR of voltage imaging in tissue. The shot noise-limited SNR for optical voltage measurements is:

$$\text{SNR} = \frac{\Delta F}{\sqrt{F+B}}, \qquad \text{[Eq. 1]}$$

where ΔF is the change in sample fluorescence due to the electrical event of interest, F is the baseline fluorescence of the indicator, and B is the background fluorescence (all measured in photon counts). There are many ways to increase the SNR. Increases in laser power increase ΔF, F, and B proportionally, leading to only square root improvements in SNR. Enhancements to GEVI brightness or expression level also increase ΔF, F, and B proportionally (assuming background fluorescence is dominated by out-of-focus reporter molecules). Thus brightness also only helps SNR by its square root.

On the other hand, enhancements to the focal plane signals, ΔF and F, without changing B, can increase SNR linearly with the enhancement in the high background limit (B >> F), which applies for one-photon GEVI measurements in tissue. Thus the 3.5-fold enhancement in focal-plane voltage signal achieved through 2P photoactivation is equivalent, from an SNR perspective, to a 14-fold enhancement in overall brightness.

The concept of photo-activation for enhanced signal localization has previously been applied in other domains of fluorescence imaging. At the single-molecule level, this phenomenon enables STORM and PALM superresolution microscopies. Multi-photon photoconversion of DsRed has been used as an optical 'highlighter'.[22] A photoactivated $Ca^{2+}$ indicator has been used for optically targeted measurements *in vivo*.[23] However, in these reporters, the photo-activation was irreversible, so imaging could only be performed on one set of neurons per sample. NovArch has the useful property of being reversibly switchable, allowing for sequential targeting of distinct cells or focal planes.

Spectroscopic studies on homologous mutations in bacteriorhodopsin (BR) provide some insights into the molecular mechanism of photoactivation in NovArch. The BR mutation V49A, analogous to V59A in Arch, has been studied in detail. In BR, this mutation stabilizes the N state, increasing its lifetime from ~5 ms to ~100 ms.[24] In our model of voltage-sensitive fluorescence in wild-type Arch, the fluorescent Q state is reached from the N state,[12] so stabilization of this photocycle intermediate would be expected to lead to photoactivation behavior. Several other mutations in BR have also been found to stabilize the N intermediate [25]. These might be plausible alternative mutations sites for making photoactivatable GEVIs.

Remarkably, a report on photochromic mutations in BR found that the mutant D85N/V49A had exceptionally high photoswitching efficiency [26]. In BR, aspartic acid 85 is the proton acceptor from the Schiff base. NovArch contains a similar mutation at the homologous position, D95Q, in which the glutamine (Q) probably plays a similar role to asparagine (N).

Based on sequence homology to bacteriorhodopsin, the V209I mutation resides in the extracellular loop between the F and G helices and the I213T mutation resides near the extracellular face of helix G, both far from the retinal chromophore (Fig. S3B). The mechanisms by which these residues affect photoactivation is unclear, though their proximity to each other suggests that saturation mutagenesis of the surrounding neighborhood would be a plausible strategy for further enhancing the photoactivation.

Optical mapping of membrane voltage in dendrites in tissue has previously been performed with fluorescent dyes introduced through a patch pipette, [27, 28] and in fly visual system *in vivo* with the GEVI ASAP2f, though these measurements required averaging over dozens of cells and many flies[29]. After averaging 1,900 spikes, the GEVI Ace2N-4AA-mNeon reported membrane voltage in dendrites immediately adjacent to the soma in mouse visual neurons *in vivo*.[5] In contrast, NovArch reports single-trial back-propagating action potentials in intact rodent brain tissue. The ability to detect single spikes with NovArch to depths of up to 200 μm in tissue suggests that this strategy may be applicable to all-optical electrophysiology measurements that penetrate the superficial neuropil to reach to cortical layer 2/3 *in vivo*.

**Acknowledgments**
This work was supported by the Howard Hughes Medical Institute, and US National Institutes of Health (NIH) grant 1-R01-EB012498-01. DB acknowledges support by a Rubicon grant from the Netherlands Organization for Scientific Research (NWO). MPC was supported by a Gordon and Betty Moore Foundation Life Science Research Foundation Postdoctoral Fellowship GBMF2550.05. Y.A. was supported by a long-term fellowship of the Human Frontier Science Program. We thank Katherine Williams for technical assistance with molecular biology and Melinda Lee for technical assistance with cell culture, animal husbandry and acute slice preparation. We thank Sami Farhi, Shan Lou and Linlin Fan for help with mouse injections and acute slice preparation.

# Materials and Methods

**Photostick screening of photoactivatable mutants of paQuasAr3**

*Mutant library preparation*
The starting gene for the mutant library was derived from a QuasAr variant with improved expression and trafficking *in vivo* compared to the previously published QuasAr2 [16]. This variant, called QuasAr3, contained enhanced trafficking sequences, the mutation K171R, and Citrine as a fluorescent expression marker.

paQuasAr3-Citrine was initially cloned into a modified FCK lentivirus vector (mFCK), in which the original CaMKII promoter was replaced by a CMV promoter and the WPRE enhancer sequence was included at the 3' end. Ampicillin and puromycin resistance genes were also included in the construct. We denote this plasmid pLenti-CMV-paQuasAr3-Citrine-Puro. This plasmid was then used as the template for error-prone PCR with the following primers, which flanked the paQuasAr3 region:

Forward: GACACCGACTCTAGAGCGCGGATCCACCATGGTAAGTATC
Reverse: TCTCGTAGCAGAACTTGTAGAATTCTTATTCATTCTCATAACAAAG

An error-prone PCR reaction was performed as follow. In a 50 µL reaction we combined 5 µL of ThermoPol Buffer (New England BioLabs), 5 µL of 2 mM dNTP (New England BioLabs), 5 µL of 10 mM dCTP and dTTP, 5 µL of 35 mM $MgSO_4$, 1.5 µL of each 20 µM primer, 0.5 µL of the plasmid template, 1 µL of 5 U Taq polymerase and 25.5 µL of nuclease-free water. The PCR condition comprised: 1 cycle of 95 °C for 3 min, 35 cycles of the following three steps, 95 °C for 25 s, 55 °C for 30 s and 72 °C for 1 min, followed by 1 cycle of 72 °C for 10 min and then maintenance at 4 °C. The error-prone PCR products were denoted Photoactivated QuasAr (paQuasAr) and cloned back into the starting plasmid to create pLenti-CMV-paQuasAr-Citrine-Puro.

*Lentivirus production*
HEK 293T cells were grown to 80% confluence in a 10 cm dish. Fresh DMEM10 medium was exchanged 1-2 hr prior to DNA transfection. A mixture of DNA plasmids (6.2 µg paQuasAr plasmid and two virus packing plasmids: 4 µg psPAX2 (Addgene ID 12260) and 1.78 µg VsVg (Addgene ID 8454)) were added into an Eppendorf tube with 0.6 mL Opti-MEM medium followed by 36 µL of 1 mg/mL PEI solution (branched polyethylenimine, average molecular weight 25,000, Sigma Aldrich, cat #408727). The mixture was vortexed briefly and incubated at room temperature for 10 min. The DNA/PEI mixture was then added drop-wise to the dish of HEK cells and incubated for 4 hrs before changing to 10 mL fresh DMEM10 medium without antibiotics. After 48 hrs, cell culture supernatant was collected and centrifuged at 500 g for 5 min to remove cells and debris. The supernatant was further filtered using a 0.45 µm filter and aliquoted for storage at -80 °C.

*paQuasAr stable HEK cell line generation*

PaQuasAr lentivirus was thawed and added at a titer of 0.001 MOI (Multiplicity of infection) in a 10 cm dish of HEK 293T cells grown to 80% confluency. 48 hrs after infection, puromycin was added to a final concentration of 2 mg/mL. Cells were selected for 14 days to stabilize the expression of paQuasAr. The stably expressing HEK cells were then frozen in liquid nitrogen for later use.

*Screening paQuasAr followed by Photostick selection*

<u>Glass-bottom dishes covalently modified with fibronectin</u>
Glass-bottom dishes (In Vitro Scientific, D35-14-1.5-N) were covalently modified with fibronectin to facilitate subsequent photochemical crosslinking of cells to the dish. Dishes were first cleaned and chemically activated by 3 min treatment in a plasma cleaner with low-pressure ambient air. The glass was aldehyde-functionalized with a 1% solution of 11-(Triethoxysilyl) undecanal (Gelest, Inc) in ethanol, which reacted for 1 hour in a nitrogen-purged glovebox. Dishes were rinsed twice with ethanol and once with nanopure water and then cured in a vacuum oven at 65 °C for 1-2 hrs.

Fibronectin (0.1 mg/mL in PBS) was added to the dishes and incubated overnight at 4 °C or at 37 °C for 2 hrs, resulting in a covalent imine bond between the surface and free primary amines on the fibronectin. Dishes were then immersed in 0.1% Tween-20 PBS for 10 min followed by rinsing three times with PBS. Completed dishes could either be seeded with cells directly or desiccated and stored at -80 °C.

Fibronectin-coated dishes were seeded with 150k HEK cells stably expressing PaQuasAr mutants for 16-24 hrs. Prior to imaging, the culture medium was exchanged for XC buffer (125 mM NaCl, 2.5 mM KCl, 2 mM $CaCl_2$, 1 mM $MgCl_2$, 10 mM HEPES, 30 mM glucose, pH 7.3).

<u>Photostick optics</u>
Photostick experiments were performed on a custom-built microscope (Fig. S2). Red illumination was provided by six 635 nm diode lasers (Dragon Lasers), each 500 mW, for a total of 3 W. The beams were combined in pairs via polarizing beamsplitters, and then directed to the sample via a custom fused silica prism placed between the objective and the sample. The beams were coupled from the prism to the sample via immersion oil, entering the sample at close to the critical angle for total internal reflection at the glass-water interface. Blue illumination was provided by a 488 nm 300 mW LED, mounted above the sample and expanded to illuminate the whole field of view.

Violet light for photocrosslinking was provided by a 407 nm 200 mW Laser (Lilly Electronics). Collimated 407 nm laser light at the back focal plane of the objective was focused at the sample to obtain a 5 μm spot. The position of the 407 nm laser focus was set by a pair of galvo mirrors (Thorlabs GVS202) located in a conjugate plane.

Imaging was performed with a large dissecting microscope objective (MV PLAPO 2x 0.5 NA; Olympus). Fluorescence was separated from illumination light using an emission filter (Semrock #Em01- R405/488/635). A 0.63x dissecting microscope objective served as the tube lens and projected an image onto a scientific CMOS camera (Hamamatsu Orca Flash 4.0).

Screening protocol

Cells were illuminated with 640 nm light (12 s, 20 W/cm$^2$), and with weak 488 nm light (4 s, 0.1 W/cm$^2$) superimposed at intervals to probe photoactivation. Prior to each experimental run, a control measurement was performed with a dish not containing cells, for calibration of background autofluorescence levels and illumination profiles.

The fluorescence intensity at each pixel depended on the following eight initially unknown parameters:

    **$I_r$**: Illumination profile of red light
    **$I_b$**: Illumination profile of blue light
    **α**: Proportionality between red light excitation and near infrared fluorescence
    **γ**: Proportionality between blue light excitation and near infrared fluorescence
    **β**: Nonlinear coupling of blue excitation to red-excited fluorescence
    **$C_{dark}$**: dark counts of camera
    **$B_r I_r$**: background sample autofluorescence upon red light illumination
    **$B_b I_b$**: background sample autofluorescence upon blue light illumination

Six fluorescence values were measured at each pixel, as follows:

| Meas. No | Cells on dish | Red ON | Blue ON | Fluorescence value |
|---|---|---|---|---|
| 1 | - | - | + | $B_b I_b + C_{dark}$ |
| 2 | - | + | - | $B_r I_r + C_{dark}$ |
| 3 | + | - | - | $C_{dark}$ |
| 4 | + | - | + | $\gamma I_b + B_b I_b + C_{dark}$ |
| 5 | + | + | - | $\alpha I_r + B_r I_r + C_{dark}$ |
| 6 | + | + | + | $\alpha I_r \beta I_b + \gamma I_b + \alpha I_r + B_b I_b + B_r I_r + C_{dark}$ |

Two additional pieces of data comprised the total optical powers in the red and blue beams. With these eight measurements, it was possible to solve for the eight variables at each pixel. The map of β as a function of position probed the nonlinear response, and the map of α probed the overall brightness. Movies were analyzed immediately after acquisition to determine whether the measured field of view contained cells with desirable photoactivation properties.

Photostick protocol

A photochemical crosslinker phenyl azide derivative, 4-fluoro-3-nitrophenyl azide (FNPA, 15 μM), was applied as the photostick reagent. Light from a 407 nm laser (15 mW) was scanned over the target cells in a small raster pattern. The illumination time was $6 \times 10^{-2}$ ms/µm$^2$ or 100-200 ms per cell depending on cell size. Subsequently, the dish was rinsed with PBS and digested with trypsin for 5 min to detach non-illuminated cells. The remaining adhered cells were detached by gently pipetting with 10 µL PBS and transferred into a PCR tube. When more than one cell was harvested, additional dilutions were used to ensure each PCR tube contained at most one cell. Each single cell was then lysed with 2.8 µL lysis buffer (40 mM DTT, 2 mM EDTA, 200 mM KOH) at 65 °C for 10 min and neutralized with 1.4 µL of neutralization buffer (400 mM HCl, 600 mM Tris pH 7.5) before conducting a PCR reaction. Q5 NEB 2x master mix was used for a PCR amplification using the conditions: 1 cycle of 95°C for 5 min; 45 cycles of the following three steps: 95°C for 30 sec, 58 °C for 45 sec and 72 °C for 1 min; followed by 1 cycle of 72 °C for 10 min and then 4 °C for ∞.

*AAV virus preparation*
NovArch-Citrine was cloned into an adeno-associated virus (AAV) vector to create AAV2/9.CAG.FLEX.NovArch-Citrine.WPRE.SVPA, for AAV production. AAV virus was produced in the Gene Transfer Vector Core at Massachusetts Eye and Ear Infirmary & Schepens Eye Research Institute (MEEI), Harvard Medical School.

*Viral injections*
All animal experiments were approved by the Harvard University Institutional Animal Care and Use Committee (IACUC). C57BL/6 mouse pups were cryo-anesthetized at P0-1 [30]. Pups were immobilized dorsal side up, facing away from the experimenter under a stereotaxic microscope. Injections were made using home-pulled micropipettes (Sutter P1000 pipette puller), mounted in a microinjection pump (World Precision Instruments Nanoliter 2010) controlled by a microsyringe pump controller (World Precision Instruments Micro4). The micropipette was positioned using a stereotaxic instrument (Stoelting Digital Mouse Sterotaxic Instrument). Pups were injected in the right hemisphere, 1 mm lateral, 1 mm anterior and 0-2 mm ventral from lambda with a virus mixture containing $8.3 \times 10^9$ GC/mL CKII(0.4)-Cre, $1.5 \times 10^{12}$ GC/mL hSyn.CheRiff-CFP and $5.8 \times 10^{12}$ GC/mL CAG.FLEX.NovArch-Citrine. 40 nL was injected at each site. Typically 9 sites were injected per pup. Pups were placed back in their home cage once they were awake.

*Brain slice preparation*
Acute brain slices were prepared from P15–P21 male and female mice. The mice were deeply anesthetized by intraperitoneal injection of 90 mg/kg ketamine and 10 mg/kg xylazine and then perfused with carbogen (95% $O_2$, 5% $CO_2$)-saturated ice-cold slicing solution with the following composition (in mM): 110 choline chloride, 2.5 KCl, 1.25 $NaH_2PO_4$, 25 $NaHCO_3$, 25 glucose, 0.5 $CaCl_2$, 7 $MgCl_2$, 11.6 Na-ascorbate, and 3.1 Na-pyruvate. Mice were then decapitated and the brains were removed into ice-cold slicing solution and then rapidly blocked for coronal sectioning at 300 µm thickness on a vibratome (Leica VT 1200S). Slices were then incubated for 45 min at 34 °C in a carbogenated artificial CSF (ACSF) with the following composition (in mM): 127 NaCl, 2.5 KCl, 1.25 $NaH_2PO_4$, 25 $NaHCO_3$, 25 glucose, 2 $CaCl_2$, and 1 $MgCl_2$. Slices could be used for 4 – 6 h after harvest. The osmolarity of all solutions was adjusted to 300 –310 mOsm and the pH was maintained at 7.3 under constant bubbling with carbogen.

*Brain slice imaging*

Brain slice imaging was performed on a home built inverted microscope (see below). The slice was immobilized in a Warner Instruments RC-27LD flow chamber using a slice anchor (harp). ACSF, perfused with carbogen as above, was flowed through the chamber at a rate of 1.5 mL/minute.

*Two-photon Microscopy*
We built a beam-scanning two-photon microscope optimized for excitation at wavelengths between 950 and 1300 nm, and for detection of fluorescence at wavelengths shorter than 775 nm. Illumination was provided by a Spectra Physics Insight DeepSee, tunable between 680 and 1300 nm, with pulses of ~120 fs at 80 MHz repetition rate. The pulse dispersion was adjusted via an internal, motorized prism pair compressor controlled by Spectra Physics software.

The beam was steered by a pair of galvo mirrors (Cambridge Technology 6215HM40B, driven by Cambridge Technologies 671215H-1HP Micromax servo driver on the X-axis and Cambridge Technologies 671215H-1 Micromax servo driver on the Y-axis). The beam then passed through an Olympus PL Scan lens and an Olympus TLUIR tube lens onto the back aperture of an Olympus water immersion XPLN25XWMP2 objective (NA 1.05). The imaging plane was selected by moving the objective with a Physik Instrumente Pifoc P725 piezoelectic nanomanipulator, driven by a Pifoc E625 driver. Fluorescence emission was separated from back-scattered excitation light via a Semrock 775 nm long-pass dichroic beam splitter (FF775-Di01-25x36). Residual laser light was rejected using a Semrock 790 nm short-pass filter (FF01-790/SP-25).

For direct two photon imaging, fluorescence was reimaged onto a Hamamatsu H10492-13 PMT. The signal was amplified and low pass filtered through an USBPGF-S1 low pass filter (Alligator technologies), with a typical cutoff of 62.5 kHz for digitization at 125 kHz. Signals were recorded using a National Instruments PCIe 6259 board.

*Camera-based imaging*
For camera-based imaging of NovArch, optical excitation of CheRiff, and initial screening of the photoactivation effect with blue light, we used 640 nm and 488 nm Coherent Obis lasers for wide-field illumination. Beams of both lasers were combined and fed through a Gooch and Housego AOM for temporal control, collimated, and passed through an iris on an x-y stage. The iris was reimaged into the sample via the fluorescence excitation pathway; fluorescence and excitation light were separated using a 664 long pass dichroic (Semrock BLP01-664R-025). The size and position of the iris provide control of the excitation spot. Fluorescence was imaged onto an Andor Ixon X3 860 EMCCD for spatially resolved imaging of 1P or 2P signals. This pathway was also used in white light illumination when applying a patch pipette to a cell.
    Camera frame rates were Fig 4: 1 kHz, Fig. 6: 500 Hz, Fig. 7: 500 Hz. In all cases, fluorescence time-traces digitally were filtered to an effective frame rate of 370 Hz. This bandwidth was selected to block some high-frequency electronic noise while still preserving spike amplitudes and waveforms.

*Co-scanning of red and 2P focal spots*
For the double focus scanning, the beam from a Coherent Obis 640 nm laser was led through a 1:1 telescope and a pair of steering mirrors and combined with the NIR-beam from the DeepSee using a FF660-Di02 dichroic mirror. The combined beam was steered along the same pathway used for conventional two photon imaging. Fluorescence was separated from the excitation light using a custom dichroic with a 25 nm transmission band around 640 nm and a longpass edge at 775 nm (Aluxa). In addition to the 790 nm shortpass filter, a 664 nm long-pass filter (Semrock BLP01-664R-025) was added in the detection path to reject the 640 nm laser light. Fluorescence was resolved on the Andor Ixon Camera via the wide-field imaging path. For typical measurements, a field of view would be selected on the camera that would allow 500-1000 Hz frame rates.

*Patch clamp recordings*
We coupled a HEKA EPC 800 patch clamp amplifier to the setup to provide simultaneous electrophysiological and optical measurements. The pipette was positioned by a Sutter MP285 micromanipulator. Whole-cell voltage clamp and current clamp signals were filtered at 3 kHz with the internal Bessel filter and digitized with a National Instruments PCIe 6259 board. The electrophysiology and optical measurements were synchronized via custom software written in LabView.

*Data analysis of electrophysiological data.*
For figure 4, action potentials were induced by current injection via patch clamp pipette. In figure 6, measured activity was spontaneous, unless otherwise indicated. For channelrhodopsin stimulated activity, we positioned an iris in the 488 nm beam, imaged the iris into the sample, and closed the iris and positioned it so that it stimulated a single soma. For figure 7, we used widefield blue light stimulation.

Data were recorded with a 1000 (Fig. 4) or 500 Hz (Fig. 6,7) camera frame rate. To convert the movie into a data trace, we averaged the pixels in the frame weighed by their response to 2P activation. Data were Fourier filtered with a 370 Hz bandwidth to remove camera-related electronic noise. For SNR calculations, we corrected the traces for photobleaching and photoactivation artefacts. We calculated spike amplitude as the difference between the peak value of the spikes and the value of the baseline in a window around the spike; window size was adapted depending on the spiking frequency. We calculated noise by computing the standard deviation in a 20-point window without spikes and minimum biological noise, and SNR by dividing spike amplitude by this standard deviation.

*NovArch fluorescence excitation and activation spectra*
We expressed NovArch in HEK 293T cells, centrifuged the cells to form a dense pellet, and performed spectroscopy in a home-built microscope system using light from a tunable supercontinuum laser (Fianium SC-450-6). This beam was combined with a Coherent Obis 640 laser to measure the photosentization spectrum, and an Omicron PhoxX 488 nm laser to measure the excitation spectrum. The cell pellet was illuminated through a 60× oil-immersion 1.45 NA objective. Fluorescence was separated from excitation light using a 660 LP dichroic and 664 LP fluorescence filter. Fluorescence from the cell pellet was imaged onto an Andor Ixon X3 860 EMCCD. The same imaging sequence as in the screening protocol above was used. To measure the excitation spectrum, 'red' was varied between 570 and 660 nm and 'blue' was kept at 488 nm; for the photoactivation spectrum, 'red' was kept at 640 nm and 'blue' was varied between 460 and 550 nm.

**Figures**

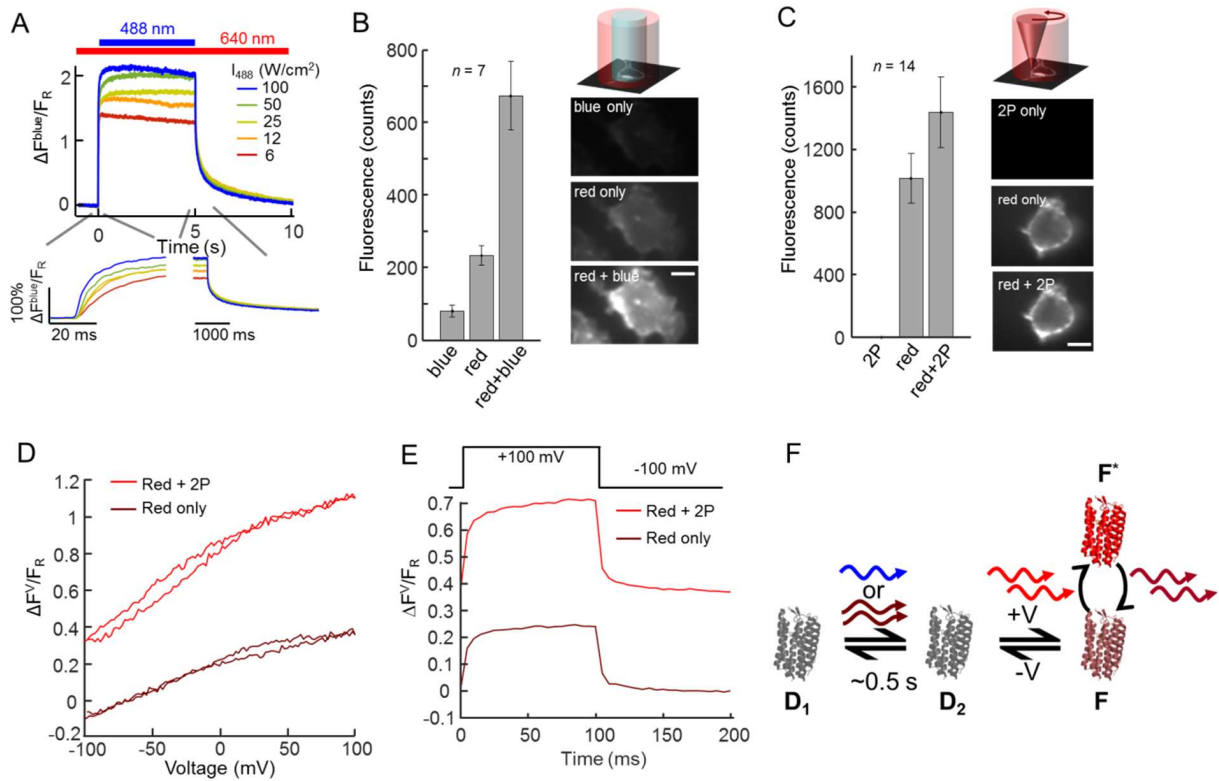

**Figure 1: Photoactivation of paQuasAr3.** A. Fluorescence response of paQuasAr3 to blue (488 nm) illumination superposed on steady red illumination (640 nm, 90 W/cm$^2$). Inset: magnified views of the fluorescence dynamics when the blue light is applied and removed. B. Fluorescence of HEK cells expressing paQuasAr3 in response to illumination with red (640 nm, 90 W/cm$^2$), blue (488 nm, 12 W/cm$^2$) and both (red + blue). Fluorescence counts are in units of counts/ms/camera pixel. Right: images of HEK cell under the three illumination conditions, shown with the same brightness and contrast scales. Scale bar 10 μm. C. Same experiment as (B), with the blue light replaced by 3 mW focused 900 nm laser light scanned along the perimeter of the cell (2P). Fluorescence counts are in units of counts/5 ms/camera pixel. D. Voltage response curve of paQuasAr3 without (brown) and with (red) 2P activation. E. Voltage step response of paQuasAr3. F. Model photocycle for photoactivation and voltage sensitivity in paQuasAr3. States $D_1$ and $D_2$ are non-fluorescent. State F can be excited by red light and emits in the near infrared.

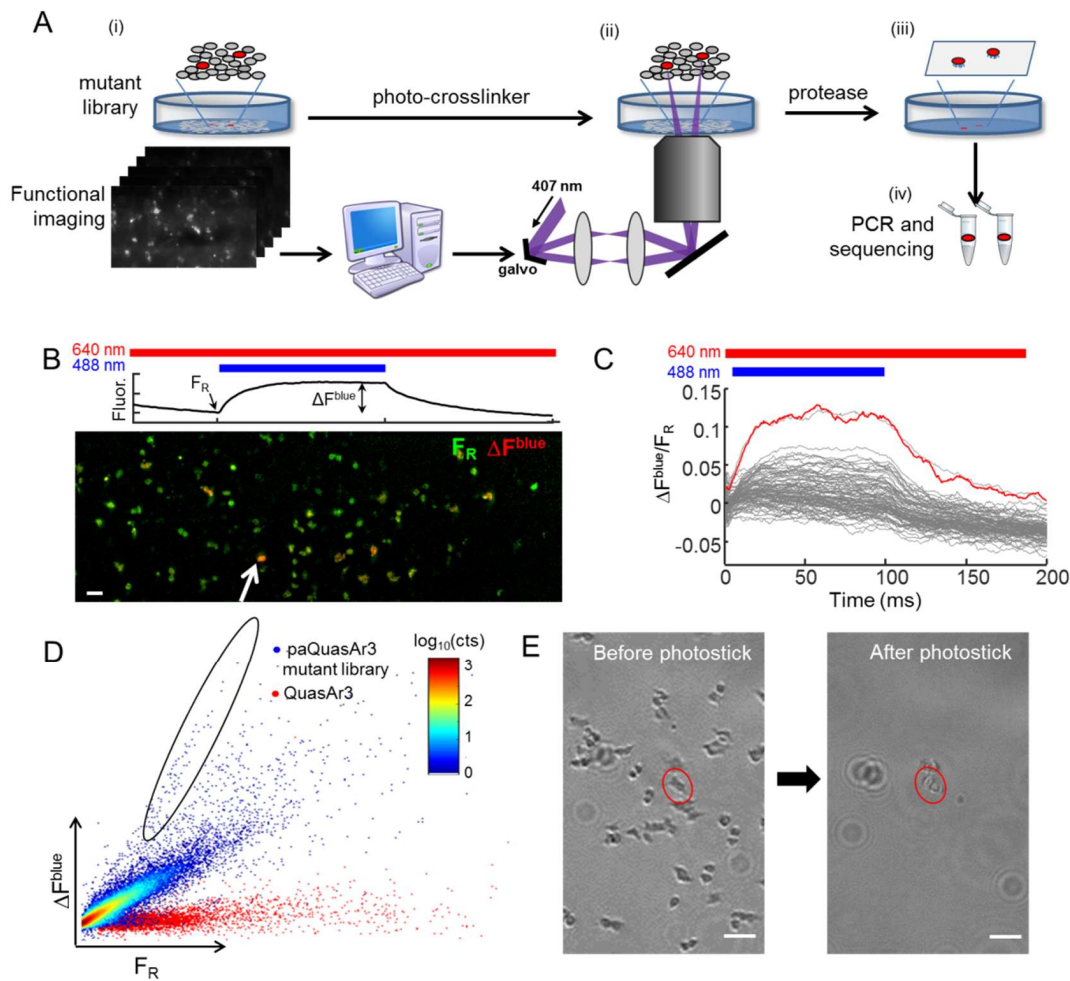

**Figure 2: Screening of NovArch.** A. Photostick screening protocol.[17] i) In a pooled library of HEK cells expressing mutants of paQuasAr3, wide-field fluorescence measurements probed blue light sensitization of red excited fluorescence. Cells with large enhancement were identified via automated image processing. ii) Focused illumination with 407 nm light crosslinked target cells to the dish. iii) Non-target cells were removed via a rinse with protease. iv) The targeted cells were detached by gentle pipetting and mutant genes were amplified and sequenced. B. Top: Illumination protocol and parameters calculated for each cell to quantify photoactivation. Bottom: Example field of view comprising HEK cells expressing mutants of paQuasAr3. Most cells expressed non-fluorescent mutants and were not visible. Green: Baseline fluorescence $F_R$ under red only illumination (640 nm, 20 W/cm$^2$). Red: Increase in fluorescence, $\Delta F^{blue}$, upon photoactivation with 488 nm light, 0.1 W/cm$^2$. Scale bar 100 μm. C. Fluorescence responses, $\Delta F^{blue}/F_R$, for individual cells in the field of view shown in (B). Red line shows response of cell indicated with white arrow in (B). D. Scatterplot of photoactivation, $\Delta F^{blue}$, and baseline brightness, $F_R$, for paQuasAr3 mutants across all fields of view and all dishes (blue dots). Control measurements with QuasAr3 showed no photoactivation (red dots). Cells with high brightness and high photoactivation were selected (circled region). E. Photostick selection of a cell expressing a targeted mutant. Left: Culture dish before photostick. Right: Same field of view after crosslinking the target cell to the dish and removing non-target cells. Scale bar 100 μm.

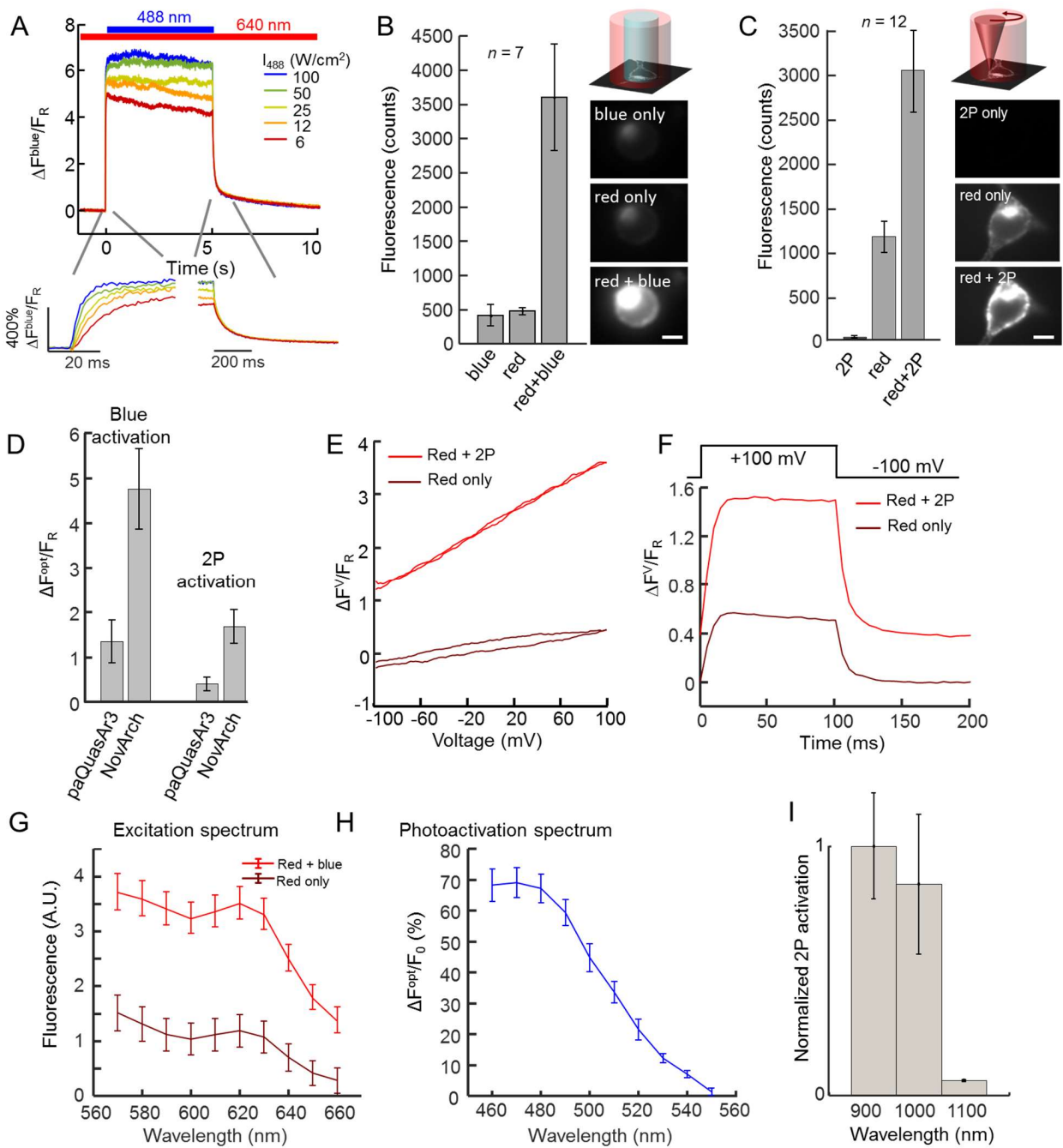

**Figure 3. NovArch is a photoactivated GEVI.** A. Fluorescence response of NovArch to blue (488 nm) illumination superposed on steady red illumination (640 nm, 90 W/cm$^2$). Inset: magnified views of the fluorescence dynamics when the blue light is applied and removed. B. Fluorescence of HEK cells expressing NovArch in response to illumination with red (640 nm, 90 W/cm$^2$), blue (488 nm, 12 W/cm$^2$) and both (red + blue). Fluorescence counts are in units of counts/ms/camera pixel. Right: images of a HEK cell under the three illumination conditions, shown with the same brightness and contrast scales. Scale bar 10 μm. C. Same experiment as (B), with the blue light replaced by 3 mW focused 900 nm laser light scanned along the perimeter of the cell (2P). Fluorescence counts are in units of counts/5 ms/camera pixel. D. Comparison of fluorescence

enhancements in paQuasAr3 and NovArch when photoactivated with blue light (488 nm, 12 W/cm$^2$) or with focused scanned 2P illumination (900 nm, 3 mW). E. Voltage response curve of NovArch without (brown) and with (red) 2P activation. F. Voltage step response of NovArch. G. Fluorescence excitation spectrum of NovArch without (brown) and with (red) blue light photoactivation. H. Single-photon photo-activation action spectrum of NovArch. I. 2P photoactivation action spectrum of NovArch.

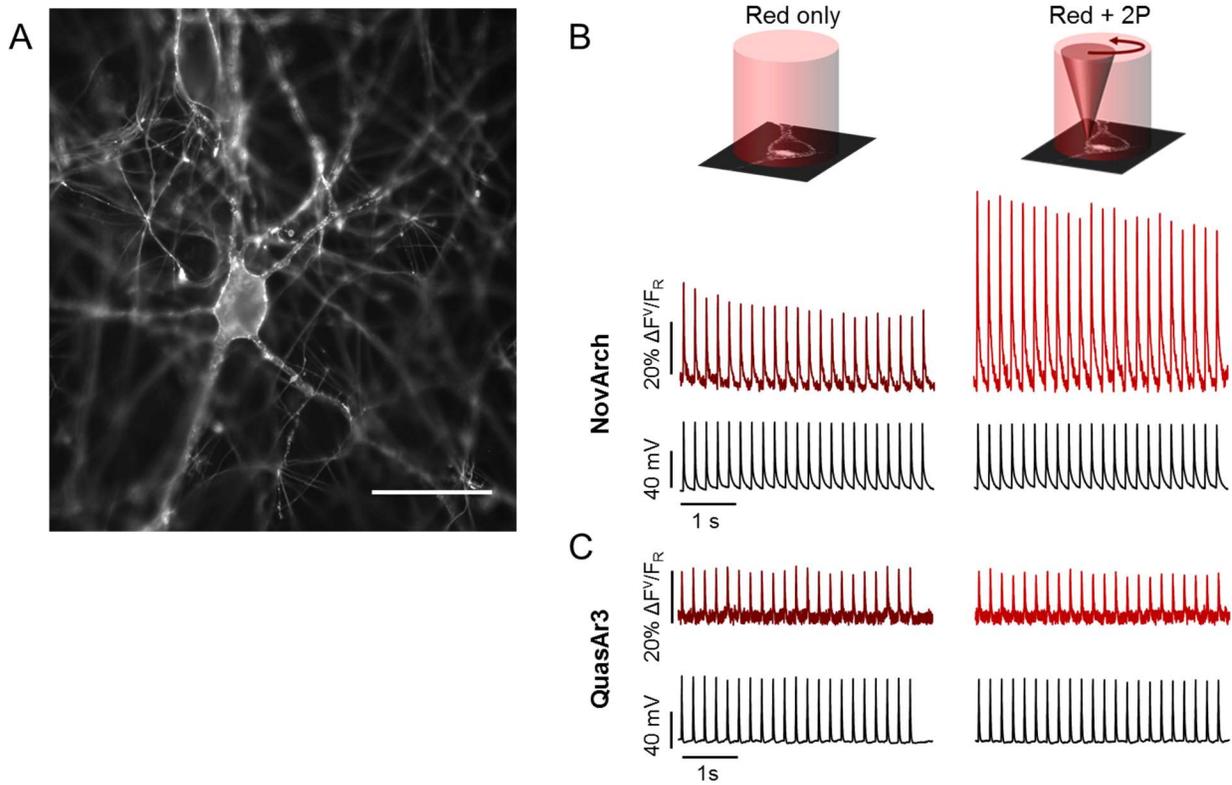

**Figure 4: 2P-activated voltage imaging in cultured rat hippocampal neurons.** A. Fluorescence of Citrine in NovArch-Citrine showed good membrane trafficking. Scale bar 50 μm. B. Simultaneous voltage and fluorescence recordings of action potentials evoked through current-clamp stimulation via patch pipette. Application of 2P illumination scanned around the cell periphery enhanced the NovArch fluorescence. C. QuasAr3 also resolved action potentials but did not show 2P photoactivation. Video was acquired at 1 kHz, digitally down-sampled to 370 Hz.

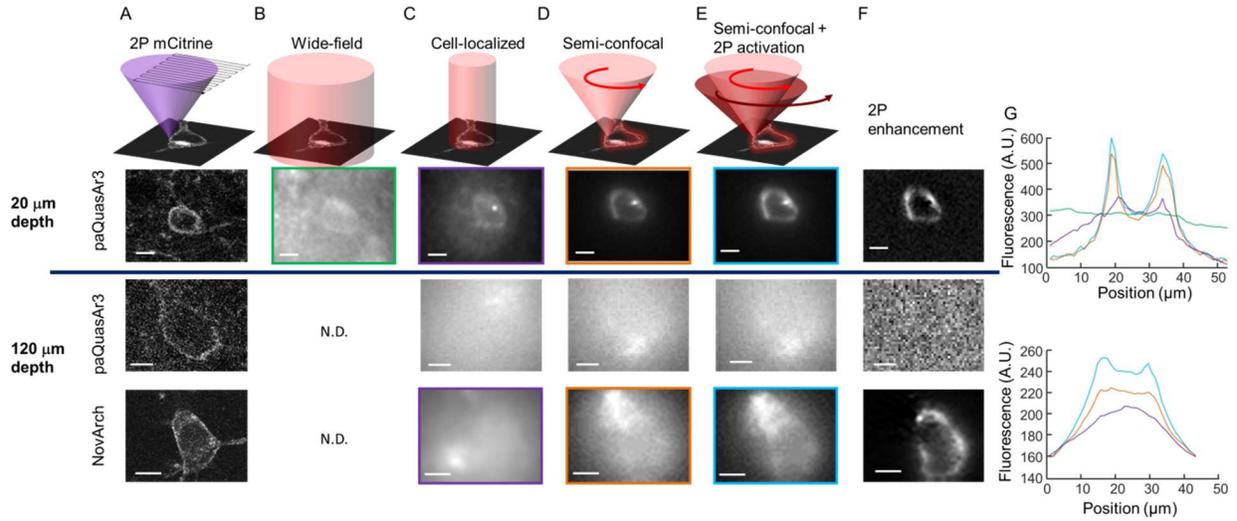

**Figure 5: Targeted illumination combined with 2P photoactivation enhance GEVI imaging in tissue.** A. Scanned 2P images of Citrine fluorescence in acute brain slices expressing the indicated GEVI fused to Citrine. Top: a near-surface cell (depth 20 μm). Bottom: cells at greater depth (120 μm). B. Wide-field illumination with red light produced a low-contrast image of fluorescence in a near-surface cell expressing paQuasAr3. No contrast was attained at a depth of 120 μm. C. Restricting illumination to the soma improved signal-to background ratio (SBR) for near-surface cells, but did not enable imaging of deeper cells. D. Scanning a focus along the equatorial periphery of the cell improved SBR for near-surface cells, and began to reveal outlines of cells at depth. E. Co-aligned 2P-photoactivation and 1P fluorescence excitation along the equatorial periphery revealed the outline of a NovArch-expressing cell at a depth of 120 μm. F. 2P enhancements calculated from the difference between the images in (E) and (D). G. Quantification of image contrast via cross sections of the cells shown in the correspondingly colored boxes in (B) – (E). All scale bars 10 μm.

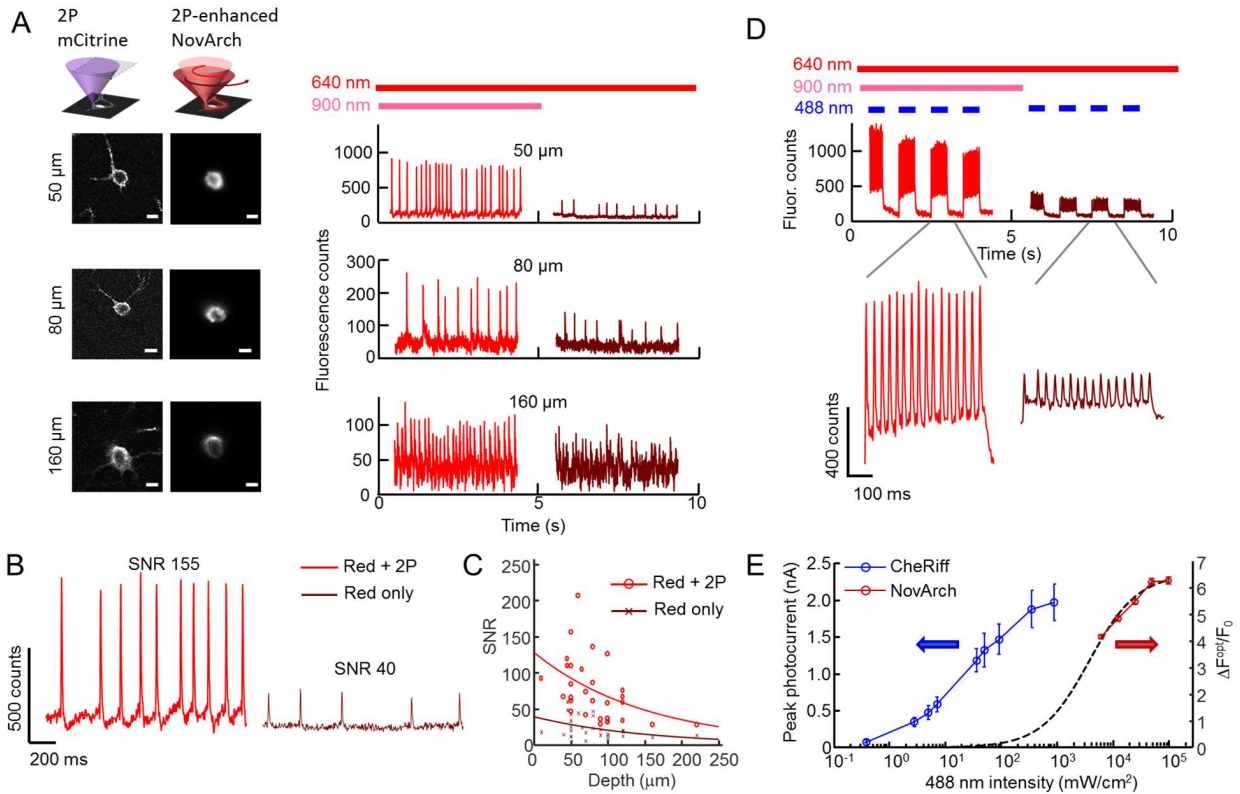

**Figure 6. Depth-resolved voltage imaging and all-optical electrophysiology in acute brain slice.** A. Left: 2P images of Citrine acquired through conventional raster scanning, and 2P-enhanced 1P fluorescence images acquired by scanning both foci along the cell periphery and acquiring fluorescence images on a camera. Scale bars 10 μm. Right: Using 2P Citrine images to define the cell periphery, 2P and 1P foci were co-aligned and scanned around neurons at different depths. The presence of 2P sensitization robustly enhanced fluorescence measurements of action potential amplitude. B. Magnified view of an optical recording from a cell at a depth of 50 μm. C. Depth-dependent SNR for single-trial AP detection, without and with 2P sensitization. The lines represent fits to an exponential decay function. D. Simultaneous optogenetic stimulation and 2P-enhanced NovArch imaging in tissue. Top: pulses of blue light excited repetitive spiking. 2P illumination enhanced spike amplitude and SNR. Bottom: magnified views of the optically induced spiking, with and without 2P enhancement. E. Comparison of the photoactivation curves for CheRiff (adapted with permission from [15]) and NovArch, both with 488 nm excitation. Blue light optogenetic stimulation at intensities between 20 – 500 mW/cm$^2$ can activate CheRiff, with little spurious sensitization of NovArch.

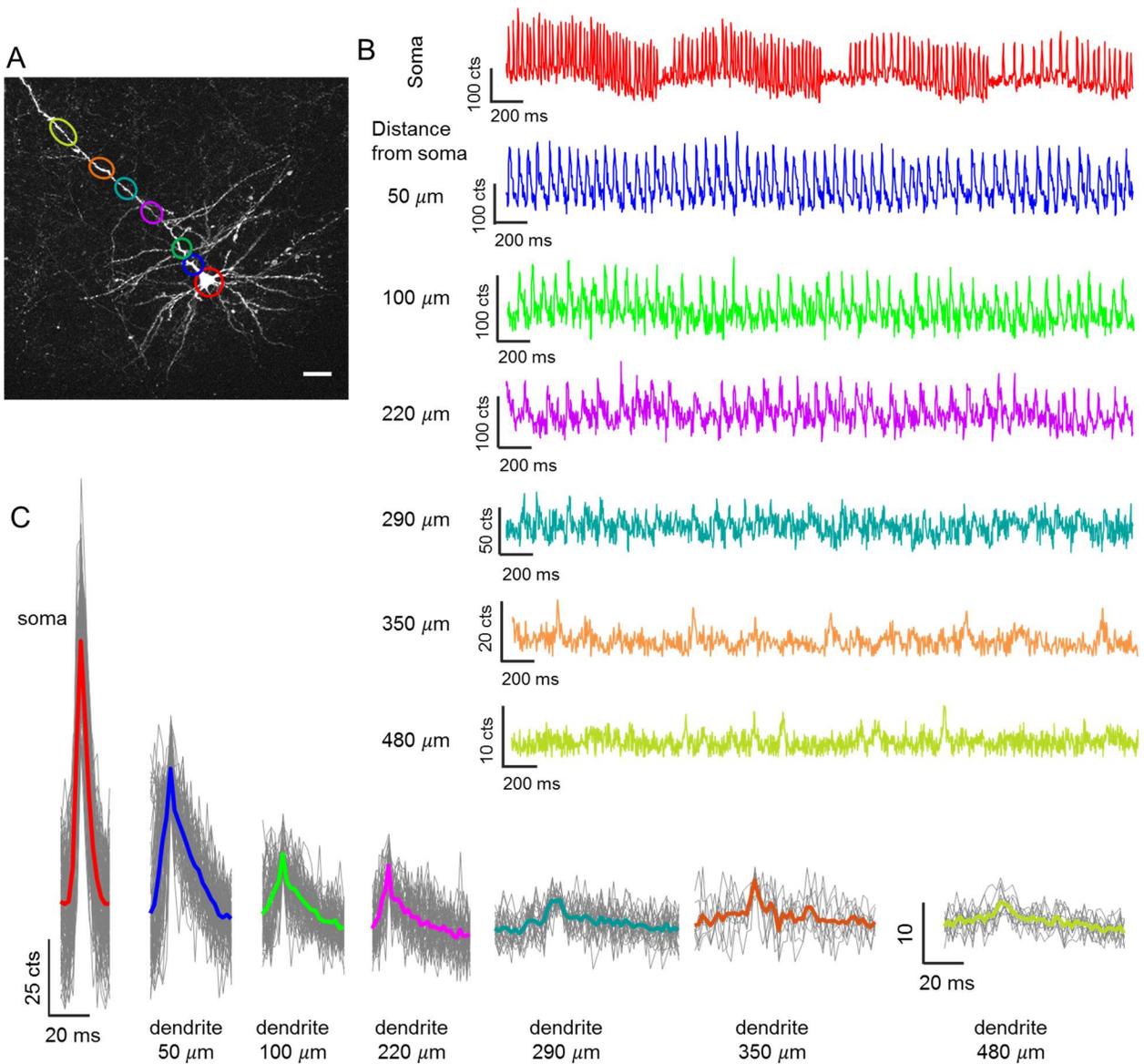

**Figure 7. Voltage imaging of dendritic back-propagating action potentials in acute brain slice.** A. Maximum intensity projection of a 2P z-stack of Citrine fluorescence in a L5 pyramidal neuron expressing NovArch-Citrine. Scale bar 50 μm. Colored ellipses indicate regions of voltage measurement. B. 2P-enhanced fluorescence recordings of action potentials at the soma and back-propagating action potentials in the dendrite, to a distance 480 μm from the soma. The periodicity in the firing in the somatic recording arises from periodic optogenetic stimulation of CheRiff. C. Spike-triggered averages of the traces in B shows the pulse shape of the back-propagating action potential as a function of distance along the dendrite.

**Supplementary Figures**

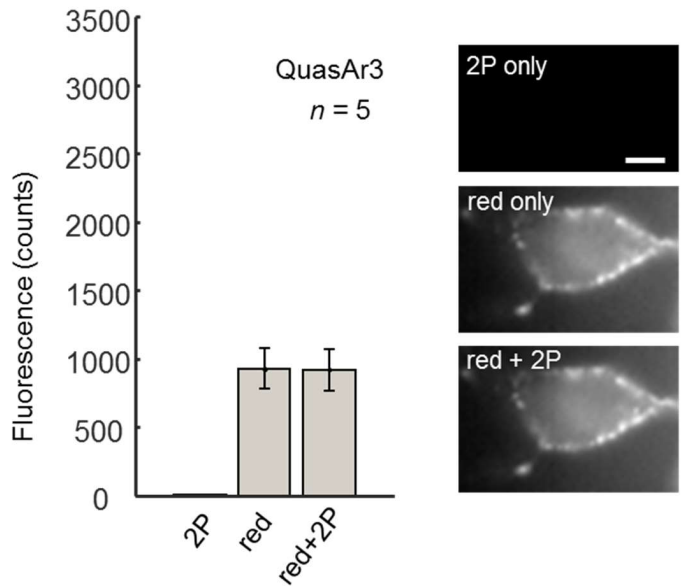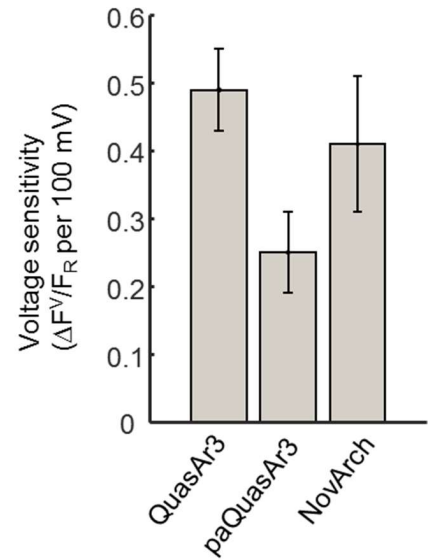

**Figure S1. Comparison of QuasAr3, paQuasAr3 and NovArch.** A. Effect of 2P illumination on fluorescence of QuasAr3. Measurements were performed as in Figs. 1C and 3C. 2P illumination did not enhance fluorescence of QuasAr3. Fluorescence counts are in units of counts/5 ms/camera pixel. Scale bar 10 μm  B. Comparison of voltage sensitivity (from -70 mV to +30 mV) of QuasAr3, paQuasAr3 and NovArch.

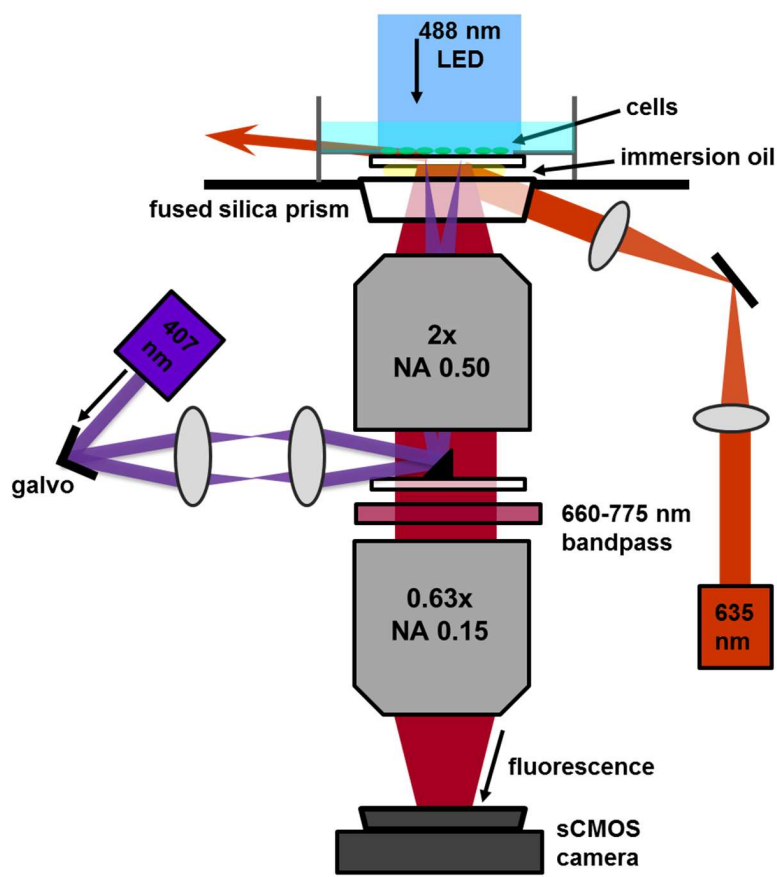

**Figure S2. Ultrawide-field microscope for NovArch screening.** High-power (3 W) red laser illumination passed through a custom fused silica prism to illuminate the sample close to the angle for total internal reflection. A 488 nm LED (300 mW) was focused onto the sample from above for homogeneous blue illumination. Fluorescence from the sample passed through a pair of high NA, large field-of-view objectives to project an image at 3x magnification onto an sCMOS camera. Violet illumination for Photostick was targeted by a pair of galvo mirrors. A small (4 × 4 mm) 45° mirror coupled the violet laser into the back aperture of the objective.

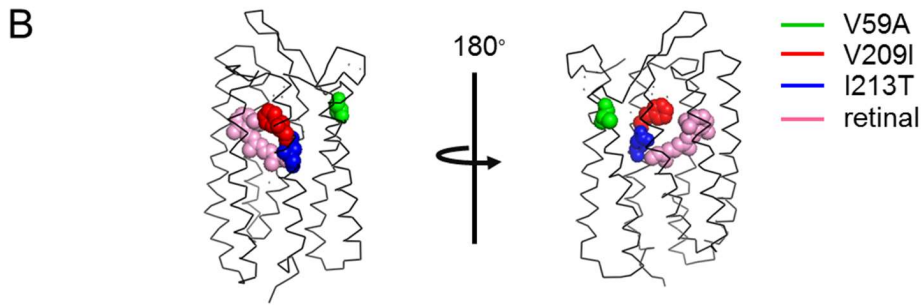

**Figure S3. Sequence comparison between NovArch and QuasAr2.** A. QuasAr2 was mutated in 4 positions to create NovArch: V59A, K171R, V209I and I213T. B. Locations of mutations important for photoactivation in NovArch, modeled on the crystal structure of Arch-2 (PDB: 2EI4): V59A (green), V209I (red) and I213T (blue). The retinal chromophore is shown in pink.

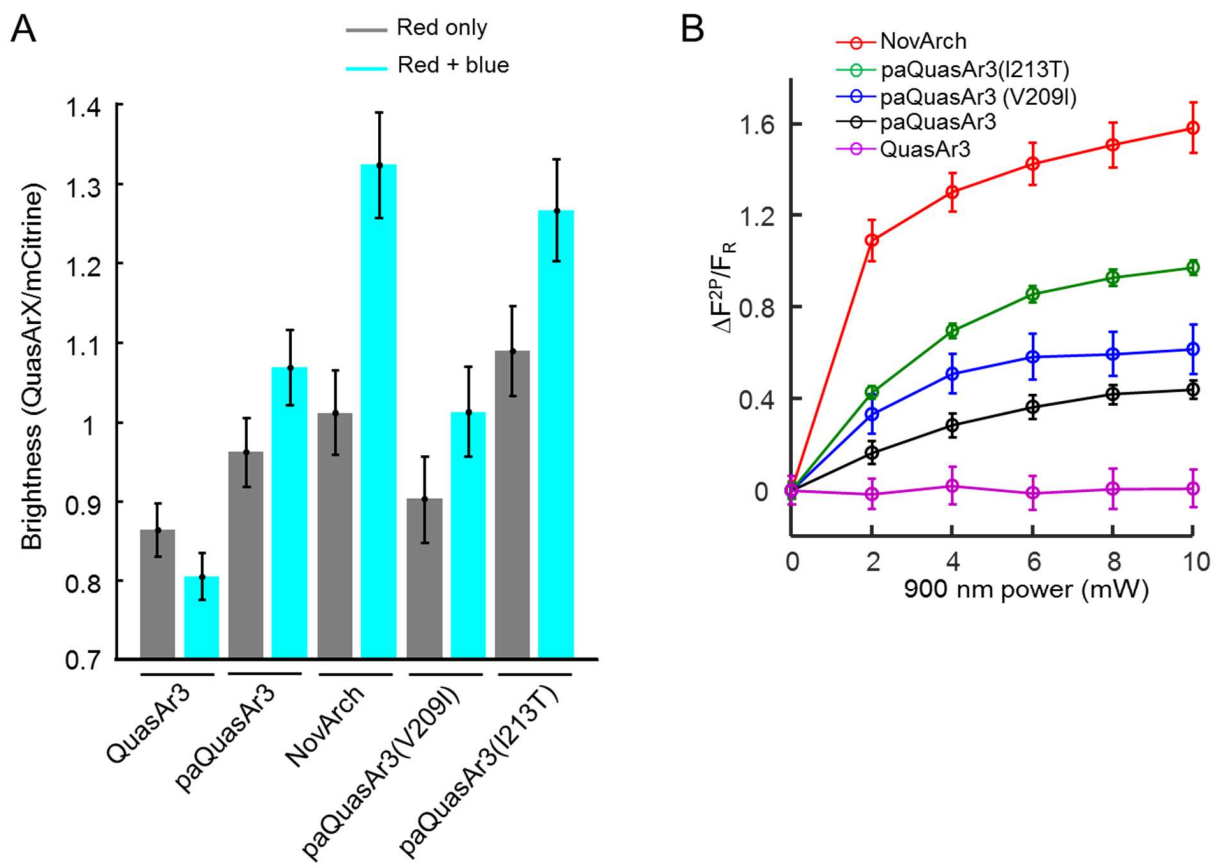

**Figure S4. Comparison of photoactivation properties of different QuasAr mutants.** A. One-photon blue light photoactivation. To compare between constructs, all QuasAr fluorescence values were normalized by the fluorescence of the appended Citrine fluorophore. Red illumination was 637 nm, 20 W/cm$^2$, blue illumination was 488 nm, 0.1 W/cm$^2$. B. Two-photon photoactivation. Both novel mutations in NovArch contributed to 2P photoactivation.

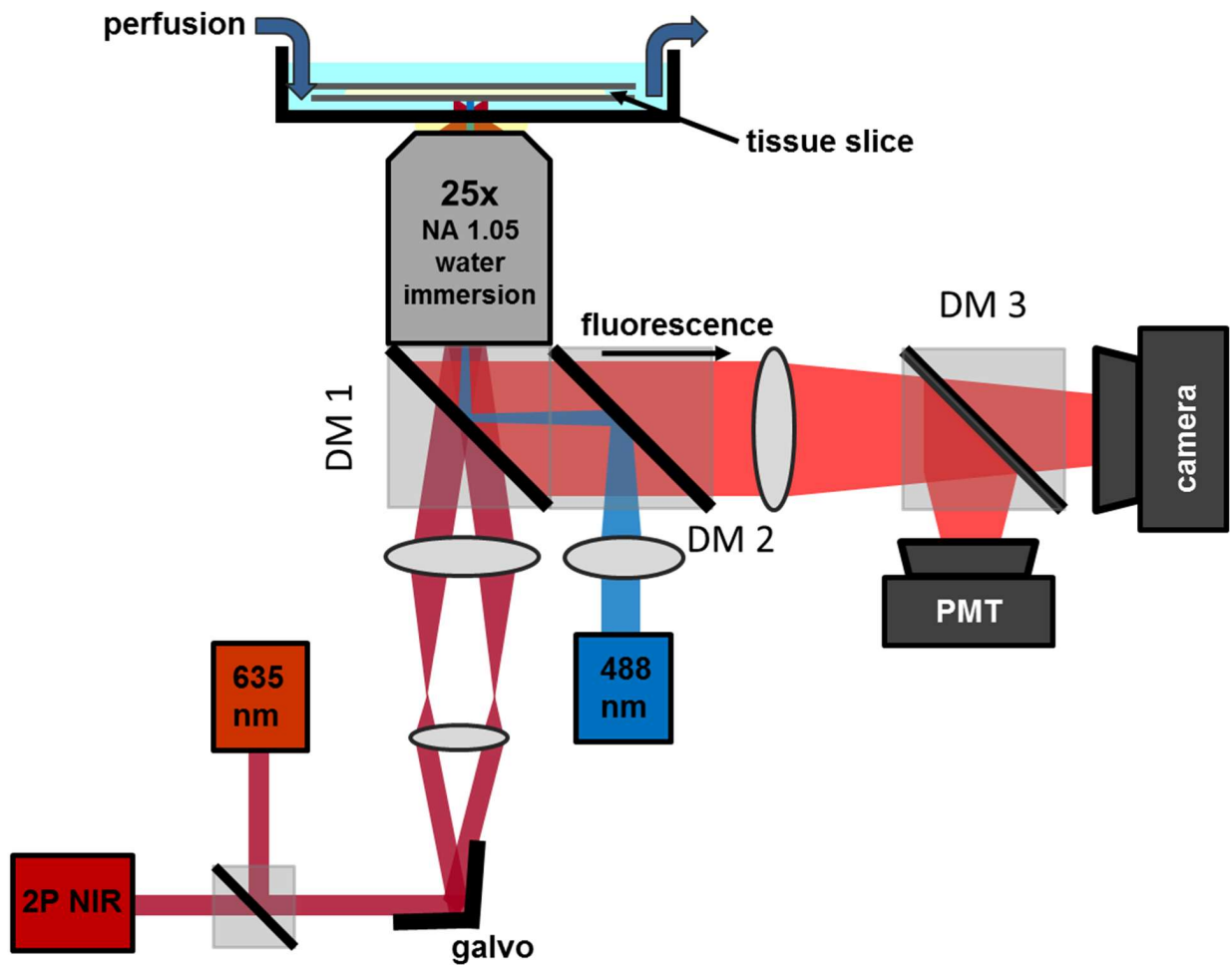

**Figure S5. Schematic of microscope for 2P-enhanced 1P fluorescence semi-confocal voltage imaging.** Beams from a 635 nm continuous wave laser and from a 900 nm femtosecond pulsed laser were combined via a long-pass dichroic mirror, and jointly steered via a pair of galvo mirrors. Fluorescence from the sample was separated from the excitation beams by a custom 640 bandpass/775 long-pass dichroic mirror (DM1). A 488 nm continuous wave beam was imaged on an aperture, which was then re-imaged into the sample plane to provide controllable wide-field blue illumination. For the comparison between wide-field and scanning imaging modalities, a second 637 nm laser beam was combined with the blue illumination pathway and DM1 was replaced with a 775 long-pass dichroic. DM2 was a quad-band dichroic separating the wide-field excitation lines from fluorescence signals. DM3 was a 660 long-pass dichroic passing NovArch fluorescence to the camera and reflecting Citrine fluorescence to a PMT for 2P Citrine imaging.

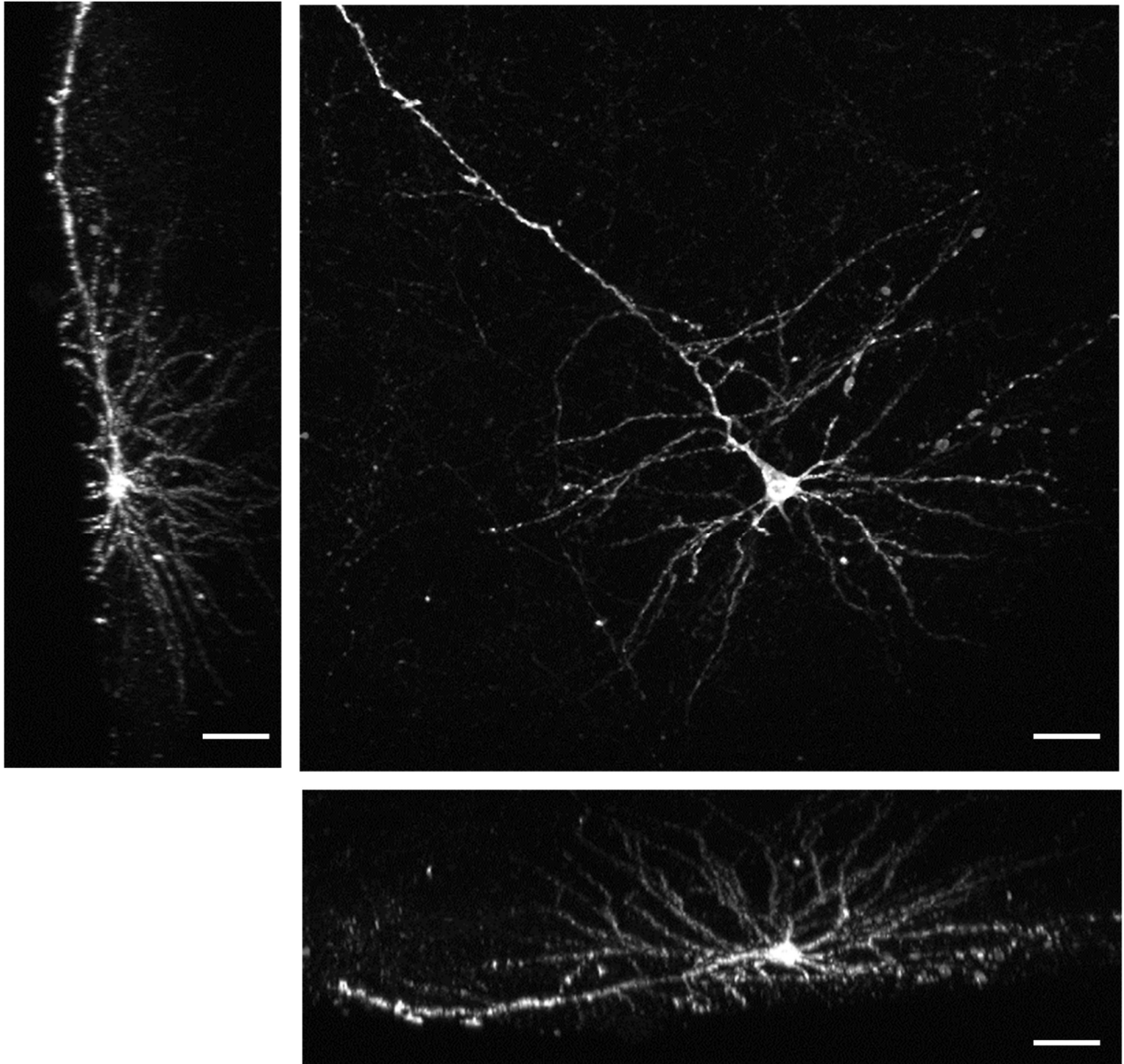

**Figure S6. Orthogonal maximum intensity projections of the cell in Fig. 7.** The complete z-stack had dimensions 600x560x200 μm. Scale bar 50 μm.